\documentclass[12pt]{iopart}

\usepackage{iopams}
\usepackage[dvips]{graphicx}
\usepackage[numbers,square,sort&compress]{natbib}

\begin{document}

\title[Hall-effect characterisation of the metamagnetic transition in FeRh]{Hall-effect characterisation of the metamagnetic transition in FeRh}

\author{M~A~de~Vries$^{1,2}$\, M~Loving$^3$, A~P~Mihai$^2$, L~H~Lewis$^3$, D~Heiman$^4$ and C~H~Marrows$^2$\eads{\mailto{m.a.devries@physics.org}, \mailto{c.h.marrows@leeds.ac.uk}}}

\address{1 School of Chemistry, The University of Edinburgh, Edinburgh EH9 3JJ, UK}

\address{2 School of Physics and Astronomy, University of Leeds, Leeds LS2 9JT, UK}

\address{3 Department of Chemical Engineering, Northeastern University, Boston, MA 02115, USA}

\address{4 Department of Physics, Northeastern University, Boston, MA 02115, USA}

\begin{abstract}
  The antiferromagnetic ground state and the metamagnetic transition to the ferromagnetic state of CsCl-ordered FeRh epilayers have been characterised using Hall and magnetoresistance measurements. On cooling into the ground state, the metamagnetic transition is found to coincide with a suppression in carrier density of at least an order of magnitude below the typical metallic level that is shown by the ferromagnetic state. The density in the antiferromagnetic state is limited by intrinsic doping from Fe/Rh substitution defects, with approximately two electrons per pair of atoms swapped, showing that the decrease in carrier density could be even larger. The surprisingly large change in carrier density is a clear quantitative indication of the extent of change at the Fermi surface at the metamagnetic transition, confirming that entropy release at the transition is of electronic origin, and hence that an electronic transition underlies the metamagnetic transition. Regarding the nature of this electronic transition, it is suggested that an orbital selective Mott transition, selective to strongly-correlated Fe $3d$ electrons, could cause the reduction in the Fermi surface and change in sign of the magnetic exchange from FM to AF on cooling.
\end{abstract}

%\date{\today}
\submitto{\NJP}

\maketitle{}

\section{Introduction}
Iron and rhodium will form an alloy at any composition~\cite{Shirane:63}, but for close-to-equiatomic compositions, from 49 to 53 atomic \% Fe, a compound of the CsCl structure with distinct mechanical, electrical and magnetic properties is the equilibrium phase~\cite{Fallot:38}. This $\alpha^\prime$-FeRh chemically ordered alloy (henceforth termed FeRh here) is an antiferromagnet (AF) that  exhibits a first-order transition to become a ferromagnet (FM) around $T_{\rm{N}} = 350 \to 400$~K~\cite{Fallot:38,Kouvel:66,Maat:05}. On heating, the metamagnetic transition between the two different but both fully ordered magnetic states (type II AF~\cite{Shirane:64} and FM) is accompanied by an isotropic 1\% volume expansion~\cite{deBergevin:61,Zakharov:64}, a large entropy release~\cite{Annaorazov:96}, and a large drop in the resistivity~\cite{Kouvel:66}. The metamagnetic transition temperature is highly sensitive to the composition $x$ in Fe$_x$Rh$_{1-x}$~\cite{Shirane:63,Driel:99} and chemical doping~\cite{Schinkel:74}, $T_{\rm{N}}$ decreases with increased applied magnetic field~\cite{Baranov:95,Maat:05} and increases with the application of pressure~\cite{Kamenev:97}.  Neutron diffraction~\cite{Bertaut:62,Shirane:64} and more recently XMCD measurements~\cite{Stamm:08} indicate that part of the 3.3$\mu_{\rm{B}}$ magnetic moment centred on the Fe in the AF phase is transferred to the Rh in the FM phase with, with $\mu _{\rm{Fe}} \sim 2.2 \mu _{\rm{B}}$ and $\mu _{\rm{Rh}} \sim 0.6 \mu _{\rm{B}}$. The Curie temperature for the high-temperature FM phase is $\sim 670$~K~\cite{Kouvel:66}, comparable to the Curie temperature of alloys with $x > 0.53$~\cite{Shirane:63}.

The transition also occurs in epitaxially-grown and polycrystalline FeRh thin films~\cite{Lommel:67, Ohtani:93, Driel:99, Maat:05}, albeit usually with a wider temperature hysteresis ($\sim 15$~K instead of the $\sim 5$~K usual for bulk specimens). Thin films of FeRh have been of much recent interest because of a potential application in heat-assisted high-density magnetic recording, taking advantage of the extremely fast switching (within a few ps) from AF to FM that can be achieved with ultrashort laser pulses~\cite{Ju:04,Thiele:03}. Further research on FeRh thin films is now focussed at tuning the transition towards specific applications, by growing epitaxially on a variety of substrates and in heterostructures~\cite{Maat:05,Ding:08,Fan:10,Loving:12}.

There is an ongoing debate about the origin of this magneto-structural and electronic transition~\cite{Derlet:12, Gray:12}. Competing explanations suggest the transition is driven either by thermal lattice expansions and lattice instabilities~\cite{Kittel:60, Moruzzi:92}, changes in the electronic density of states at the Fermi surface~\cite{Tu:69, Annaorazov:96, Kobayashi:01} and changes in the magnetic entropy due to an increase in magnon modes in the FM state with respect to the AF state~\cite{Gu:05}. In recent work, using an empirical model, it was suggested the transition arises from entropy changes related to both volume-per-atom and magnetic moment degrees of freedom~\cite{Derlet:12}.

Recent hard X-ray photoemission measurements~\cite{Gray:12}, which are sensitive to the bulk of the thin film, have revealed large quantitative as well as qualitative changes in the electronic
structure at the AF-FM transition. This is in line with heat capacity measurements that reveal a clear change of the Sommerfeld coefficient at the transition, signifying a change in the density of states at the Fermi level~\cite{Tu:69, Annaorazov:96, Kobayashi:01, Gray:12}. In this case a large change in the carrier density might also be expected at the transition. Hall-effect measurements, with which such a change in carrier density could be quantified, could therefore decide whether the entropy release, and hence the origin of the transition, is electronic or not. Surprisingly, Hall-effect data are presently only available for the FM state in Ni-doped FeRh. Here, we report further characterisation of the change in the electronic properties at the metamagnetic transition, quantifying the change in carrier density at the transition using Hall-effect and magnetoresistance (MR) measurements on FeRh epilayers. We show that the carrier density in the AF phase is anomalously low for a metal, and that it depends on the degree of Fe$\Leftrightarrow$Rh antisite disorder, with each pair of interchanged atoms yielding two electrons as current carriers.

\section{Experimental}
\begin{figure}
  \begin{indented}\item[]
  \includegraphics[width=10cm]{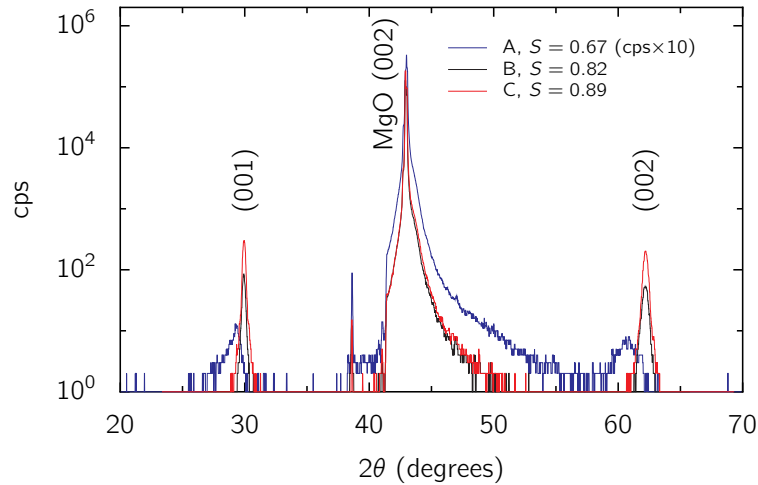}
  \end{indented}
  \caption
    {(Color online) The room temperature specular X-ray diffraction spectra of three samples with different degrees of substitutional disorder (with order parameter $S$). \label{fig:xrd}}
\end{figure}

The samples we studied were (001)-oriented epilayers of FeRh, of thickness $t$ in the range 30-50 nm, grown on (001) MgO substrates. They were grown using dc sputter deposition from a single
Fe$_{47}$Rh$_{53}$ target at 870~K, at a rate of $\sim 0.3$~\AA/s. The FeRh film composition differs slightly from that of the sputter target, and is closer to 50:50 at\%.  Prior to deposition, the MgO substrates were heated overnight at 970~K, with the vacuum in the chamber rising to $3\times 10^{-6}$~Torr. The background pressure during growth was typically $3\times10^{-7}$~Torr. Hence, Ar with 4\% H$_2$ was used as sputter gas to avoid sample oxidation during growth. Samples with different degrees of substitutional disorder, expressed in the order parameter $S = r_{\rm{\alpha}}+r_{\rm{\beta}}-1$ (where $r_{\rm{\alpha}}$ ($r_{\rm{\beta}}$) is the ratio of the Fe (Rh) located on the Fe (Rh) site), were obtained by subsequent in-vacuum annealing at temperatures $T_{\rm{A}}$ up to 1070 K. Chemically disordered FeRh is effectively bcc, for which the (001) Bragg reflection is forbidden, while fully CsCl-ordered FeRh is cubic primitive, for which the (001) reflection is allowed. The order parameter $S$ is thus obtained from X-ray diffraction using $S = \sqrt{I^{\rm{exp}}_{001}/I^{\rm{exp}}_{002}} / \sqrt{I^{\rm{calc}}_{001}/I^{\rm{calc}}_{002}} \approx
\sqrt{I^{\rm{exp}}_{001}/I^{\rm{exp}}_{002}}/1.07$, where $I^{\rm{exp}}_{00l}$ and $I^{\rm{calc}}_{00l}$ are the experimental and theoretical intensities of the (00$l$) Bragg reflection, respectively~\cite{warren_x-ray_1969}. For the calculation of the theoretical intensities the Debye-Waller factors from extended X-ray absorption fine structure measurements on FeRh were used~\cite{Miyanaga:09}. Fig.~\ref{fig:xrd} shows the specular X-ray diffraction patterns of samples A, B and C, typical for an as-grown sample, a film annealed at 970~K, and a film annealed at 1070 K, respectively. At lower annealing temperatures the intensity of the (002) Bragg reflection is also reduced, which is particularly clear in Fig.~\ref{fig:xrd} for the as-grown sample. It appears that, at least to some extent and for our growth method, epitaxy and Fe/Rh sublattice ordering are associated.

Hall bars were patterned from the FeRh films using ion-beam milling so that the longitudinal and Hall resistivities could be measured simultaneously, with dc and low-frequency ac currents of $\sim 1$~mA. Four additional samples, denoted C', D, E and F were studied in less detail. Details of the samples used are listed in Table~\ref{table:samples}. Hall measurements were carried out in an 8~T Oxford Instruments magnet at temperatures between 2 and 280 K, and in an open-bore 14~T cryogen-free magnet (Cryogenic Limited CFM-14 T-50) at temperatures between 293 and 377~K. Fig.~\ref{fig:transition}(d) shows the field-temperature dependence of the transition for sample B in this region. The field was scanned from zero to $\pm 14$~T at constant temperatures between 293 and 377 K, accessing the FM phase at progressively lower fields as the temperature is increased.

\begin{table}[b]
  \caption{Sample details. $t$ is the sample thickness. $T_{\rm{A}}$ is the approximate post-growth annealing temperature. $S$ is the order parameter determined from X-ray diffraction as described in the text. The residual resistivity ratio (RRR) is defined as the ratio of resistivities at helium and room temperatures. $T_{\rm{N}}^{\uparrow}$ and $T_{\rm{N}}^{\downarrow}$ are the midpoints of the transition on heating and cooling, respectively. Sample C is the same film as  C$^\prime$, annealed at a higher temperature after initial measurements. The carrier densities in the AF phase obtained from all  these samples are shown in Fig.~\ref{fig:nvsS}. }
  \label{table:samples}
  \begin{indented}
  \item[]
  \begin{tabular} {@{}lllllll}
  \br
  Sample & $t \pm 2$(nm) &  $T_{\rm{A}}$ (K) & $S\pm0.03$ & RRR &
  $T_{\rm{N}}^{\uparrow}$ (K) & $T_{\rm{N}}^{\downarrow}$ (K) \\
  \mr
  A & 40 & 870 & 0.67 & 1.23& N/A & N/A\\
  B & 50 & 970 & 0.82 & 3.6 & 370 & 343\\
  C & 40 & 1070& 0.89 & 5.5 & 400 & 379\\
  C$^\prime$& 40 & 970 &  0.84 &-- & 363 & 341\\
  D & 40 & 1070& 0.90 & 5.25& 408 & 390\\
  E & 40 & 1000& 0.88 & --  & -- &-- \\
  F & 30 & 970 & 0.85 & --  & -- &-- \\
  \br
  \end{tabular}
  \end{indented}
\end{table}

%\begin{figure}
%\includegraphics[width=7cm]{mount_phasedia.eps}
%\caption{(Color online) (a) A mounted sample with patterned Hall bars. The chip is an 8 mm square. (b) A schematic field-temperature phase diagram around the first-order AF-FM transition in sample B. The lower and upper (blue and red) diagonal lines mark the midpoints of the transitions on cooling and heating, respectively. The gradient indicates the hysteretic regime. The horizontal lines indicate the scanning range in Fig.~\ref{fig:transition}, at $T = $ 293, 314, 334, 354, and 377~K.}
%\label{fig:pd}
%\end{figure}

\section{Results}
\begin{figure}
  \begin{indented}\item[]
  \includegraphics[width=12cm]{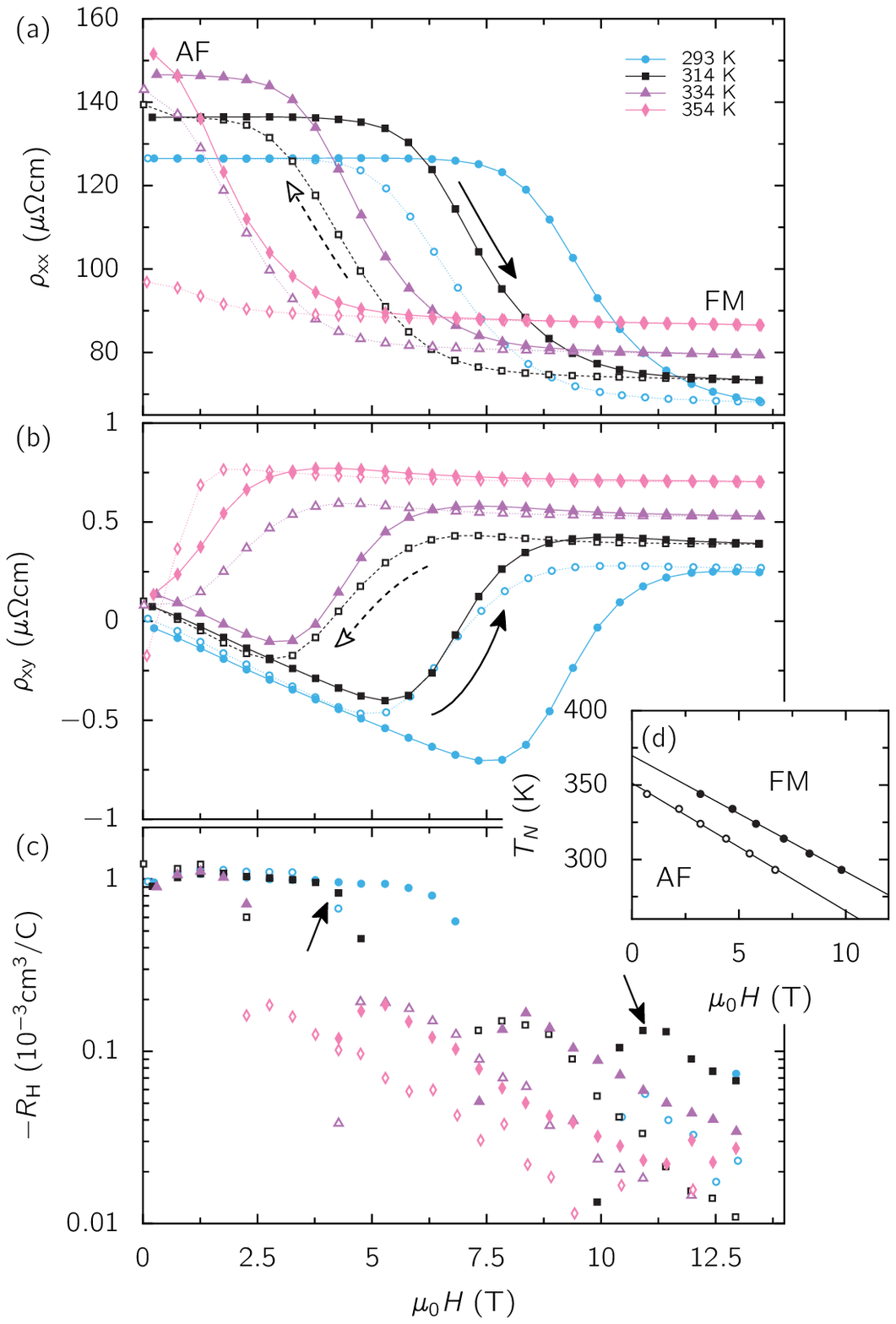}
  \end{indented}
  \caption
  {(Color online) The longitudinal resistivity $\rho _{\rm{xx}}$ (a), the Hall resistivity $\rho_{\rm{xy}}$ (b) and the Hall coefficient $R_{\rm{H}}$ (c) of sample B across the AF-FM transition induced by magnetic fields up to $\sim14$~T. The closed and open symbols correspond to the data
    taken in increasing and decreasing magnetic fields, respectively, revealing the magneto-electronic hysteresis. (d, inset) The measured $H,T$ phase diagram with the midpoints of the transitions on heating (closed symbols) and cooling (open symbols). \label{fig:transition}}
\end{figure}

Fig.~\ref{fig:transition}(a) shows the field dependence of the longitudinal resistivity ($\rho _{\rm{xx}}$) of sample B. There are three characteristic features in these data: the transition is clearly observable as a drop in $\rho _{\rm{xx}}$ by a factor of $\sim 2$ as usual; there is a sizable hysteresis for temperature and magnetic fields sweeping up and down as expected for a first order transition; and the transition moves to lower fields at higher temperatures at the usual rate of about 8~K/T. These are the characteristics typically found in FeRh.

Fig.~\ref{fig:transition}(b) shows the simultaneously measured Hall resistivity $\rho_{\rm{xy}}(H) = V_{\rm{xy}}t/I_{\rm{xx}}$. From the $\rho _{\rm{xy}}(H)$ data taken at 293, 314 and 334 K it is clear that in the low-field AF phase the Hall resistivity varies linearly with field, with a large negative Hall coefficient. This arises from the ordinary Hall effect, with transport dominated by $n = 0.19$ electron-like carriers per unit cell and a room temperature mobility $\mu \sim 7$~cm$^2$/Vs for sample B. This carrier density is anomalously low, about an order of magnitude smaller than is usual for transition metals. No change in the carrier density could be detected within the AF state down to 100 K. Below this temperature the carrier density gradually falls, with at 2~K 70\% of the value at 100~K.

In contrast, in the FM phase $\rho _{\rm{xy}}$ is markedly non-linear in $H$. The sharp increase of $\rho _{\rm{xy}}$ with increasing field at the edges of the hysteretic transition region is
attributed to onset of the anomalous Hall effect (AHE) as the sample becomes FM and the moment is pulled out-of-plane in the applied field. Beyond that, $\rho _{\rm{xy}}$ continues to vary its slope with respect to field, asymptotically approaching a value constant in $H$. Fig.~\ref{fig:transition}(c) shows the field dependence of the Hall coefficient $R_{\rm{H}} = \mathrm{d}\rho_{\rm{xy}}/\mathrm{d}B$. From DC susceptibility data in the FM state and in large fields it was confirmed that the out-of-plane moment saturates at 2~T, indicating that changes in the Hall coefficient in larger fields are not due to the anomalous Hall effect. It is clear that the magnitude of the Hall coefficient drops by nearly an order of magnitude from $-R_{\rm{H}} \sim 10^{-3}$ to $10^{-4} $cm$^{3}$/C across the transition, shown by two arrows for the 314 K data (the rapid varation of $R_{\rm{H}}$ within the hysteretic regime is not shown for clarity). In the FM phase $R_{\rm{H}}$ becomes smaller by at least another order of magnitude and further decreases with increasing field. This dramatic change in $R_{\rm{H}}$ on fully entering the FM phase, to values that are typical for transition metals, is much larger than the relatively small changes observed at the onset of charge-density/spin-density wave AF order in bcc Cr\cite{kummamuru:08}.

\begin{figure}
  \begin{indented}\item[]
  \includegraphics[width=10cm]{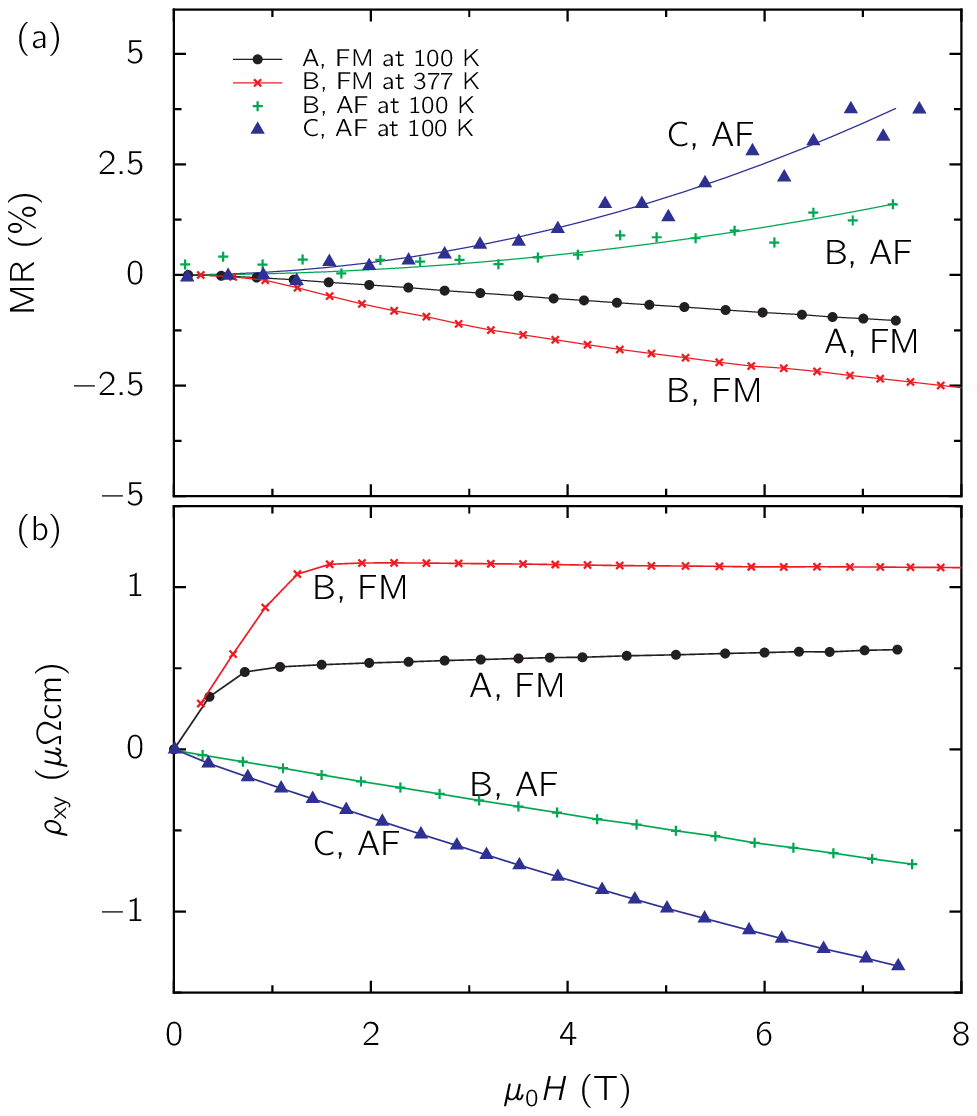}
  \end{indented}
  \caption
    {(Color online) The magnetoresistance (MR) (a) and the Hall resistivity $\rho_{\rm{xy}}$ (b) in the ferromagnetic sample A, in the AF and FM states for sample B and in the AF state for sample C. \label{fig:comp}}
\end{figure}

While a single $R_{\rm{H}}$ is obtained for all the data in the AF phase on this sample, in the FM phase the particular field dependence of $R_{\rm{H}}$ depends on the distance from the AF-FM transition, rather than the actual field; at each temperature measured, $R_{\rm{H}}(\mu _0H)$ is shifted along the field axis. Fitting the measured $R_{\rm{H}}(\mu _0H)$ with the usual two-band model~\cite{Hurd:72} will in this case clearly lead to arbitrary, temperature dependent results and we conclude that the only information we can take from this data is that in the FM phase $R_{\rm{H}} \to 0$ as $H$ increases. Within the two-band model~\cite{Hurd:72}, a vanishing $R_{\rm{H}}$ points to two types of carrier, with mobilities that are opposite but equal in magnitude, i.e. $n_{\rm n} = n_{\rm p} = n$ and $\mu_{\rm n} = - \mu_{\rm p} = \mu$. The MR will then be $\rho _{xx}(B) = (n e \mu )^{-1}(1+\mu^2 B^2)$, so that the mobility and carrier density can in ideal cases still be obtained. The MR for sample B in the FM phase at 377 K is shown in Fig.~\ref{fig:comp}(a) alongside the typical MR for sample B in the AF phase at 100~K, sample A (with an FM ground state), and sample C (AF at 100~K). It is interesting that the MR is negative in the FM state of sample B, as for the FM sample A, as usual due to the suppression of
magnon-scattering in a magnetic field~\cite{Mihai:08}. Hence the two-band model can in this case not be used to extract the actual carrier densities and mobilities in the FM phase. It is clear, however, that in the FM regime transport is dominated by these newly released carriers, implying an order of magnitude increase in carrier density at least, to values of $\sim 1$ carrier per unit cell, as is normal for metals. In Fig.~\ref{fig:comp}(b) the Hall resistivities of samples A, B, and C are compared. Neglecting the steep rise in the Hall resistivity up to $\sim 1$~T for sample A (and B) due to the AHE, $\rho _{\rm{xy}}$ increases linearly for sample A with $R_{\rm{H}}= 1.4\times 10^{-4}$~cm$^3$/C corresponding to a single type of carrier, with $n=(eR_{\rm{H}})^{-1} = 1.25$ hole-like carriers per unit cell. This is in contrast to the high-field gradient in $\rho_{\rm{xy}}$ for sample B in the high-temperature FM state, which is almost vanishing. The carrier density for FM sample A is comparable to that of bcc Fe, with approximately 2 hole-like carriers per Fe\cite{Cottam:68}, and is independent of temperature between 4 and 250~K.

\begin{figure}
  \begin{indented}\item[]
  \includegraphics[width=8cm]{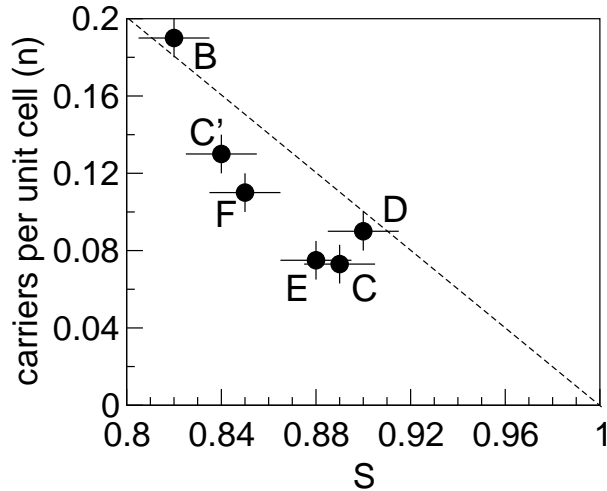}
  \end{indented}
  \caption{The carrier density $n$ as a function of the Fe/Rh sublattice order parameter $S$ for each of the six samples of FeRh. The dashed line shows a 1:1 negative correlation between these quantities as a guide to the eye.}
\label{fig:nvsS}
\end{figure}

Not only is the carrier density in the AF phase anomalously low, but it is found to vary with the concentration of antisite substitution defects. For sample C in the antiferromagnetic state $|\mathrm{d}\rho _{\rm{xy}}/\mathrm{d}B|$ is roughly twice as large as for sample B, with $R_{\rm{H}} (5~\rm{T})= -1.7 \times 10^{-3}$~cm$^3$~C and between 0.07 and 0.11 electron-like carriers per unit cell. The variation in the slope of $\rho _{\rm{xy}}(B)$ in Fig.~\ref{fig:comp}(b) is slightly greater for sample C than for sample B (where it is almost perfectly constant), and the MR of sample C is twice as large as for sample B (Fig.~\ref{fig:comp}(a)). The variation in the gradient of $\rho _{\rm{xy}}(B)$ is ascribed to a small concentration of secondary carriers, which given the reduced number of majority carriers now make up a larger fraction. Given that the Fe/Rh substitution defect concentration in sample C is $\sim 60$\% of that of sample B (Table~\ref{table:samples}), a natural explanation for the large difference in carrier densities measured in the AF phase is that the majority carriers in this phase arise from the native doping due to Fe/Rh antisite substitution. In Fig.~\ref{fig:nvsS} the Hall resistivity is shown to be correlated with the order parameter $S$ in data from these and four additional samples. The trend revealed is fairly close to two electron-like carriers per pair of atoms substituted, and confirms that the majority of carriers in the AF phase arise from native doping due to substitutional disorder. While samples with varying thicknesses were used (see Table~\ref{table:samples}), no thickness dependence was observed and so we assume that the observations reported here are likely to be equally applicable to bulk samples.

No dependence of carrier density on sample disorder was observed in the high-temperature FM phase; the data in Fig.~\ref{fig:transition} for sample B are representative of the Hall response in the high-temperature phase of FeRh.

\section{Discussion}
The dramatic changes in the Hall coefficient observed at the transition, revealing a sub-metallic carrier density in the AF phase, leave little doubt that most of the entropy release at the transition arises from an increase in carriers, tied to an increase in the density-of-states at the Fermi level. Because phase transitions are driven by an increase in entropy, this firmly establishes that, as suggested earlier~\cite{Tu:69, Gray:12}, the transition is driven by a profound change in the electronic structure.  It should be noted that the change in electronic structure can not be explained as a Slater-type transition~\cite{Koenig:82, Kobayashi:01}. In the Slater transition part of the Fermi surface is gapped out with the doubling of the Brillouin zone when already antiferromagnetically coupled spins order in a long-range-ordered antiferromagnetic N\'eel state on cooling. In FeRh the high-temperature phase is ferromagnetic, leaving the appearance of antiferromagnetism itself unexplained in a Slater-type scenario. As this Hall-effect data confirms, it is electronic ordering of some kind on cooling through the transition that gives rise to the change in magnetic exchange from ferromagnetic to antiferromagnetic.

That the volume expansion at the transition on heating is only a secondary effect is also brought out from other recent experiments on thin films. FeRh grown epitaxially on MgO has a tetragonal unit cell and expansion of the cell at the transition is only along the axis perpendicular to the film surface~\cite{Kim:09}. If the transition arose as a result of a volume instability it would be very unlikely that it persists in a system where the (isotropic in bulk) volume expansion is restricted to one dimension only. Furthermore, that the FM state can be induced within a few ps with a short laser pulse, well before the thermal energy from the laser pulse can dissipate to the crystal lattice~\cite{Thiele:03,Ju:04}, implies that the transition is not driven by a structural instability.

One could then speculate about the nature of the electronic transition and our Hall data provide a few further hints in this respect. A conductivity dominated by defect carriers in the low-temperature state with a well-defined transition to a metallic-like state on heating, as observed here for FeRh, is reminiscent of the case of NiS. This compound has a metal to non-metal transition accompanied by a magnetic ordering transition at 265~K~\cite{Sparks:68}. Hall measurements revealed that in the non-metal phase the number of carriers scales with the number of Ni vacancies in the lattice, with two hole-like carriers per Ni vacancy~\cite{Ohtani:70,Barthelemy:76}. There is of course one crucial difference; for NiS it is clear that removal of a charge-neutral Ni-atom from a Ni$^{2+}$ site creates two hole-like carriers. How in FeRh carriers arise due to substitution defects while the stoichiometry of the compound is unchanged is not \textit{a priori} clear and deserves further attention. The fact that in NiS a Mott-like transition occurs~\cite{Imada:98} despite the presence of carriers, resulting in a semimetal low-temperature state, suggests that a similar possibility might be open for FeRh. In
general, a Mott transition~\cite{Mott:61,Imada:98} can occur in materials with a half-filled $d$ or $f$ band that would according to band-theory be metals; the strong Coulomb-repulsion between the $d$ or $f$ electrons leads to a splitting of this band into two Mott-Hubbard bands. These bands are separated by a Mott-Hubbard energy gap, causing insulating behaviour with localized unpaired spins if the lower band is full and the upper band is empty. The unpaired localised spins in Mott insulators are generally coupled antiferromagnetically~\cite{Kanamori:59, Imada:98}.

As indicated by the near-linear field dependence of the Hall resistivity in the AF phase and the correlation with substitutional disorder the carrier density in a disorder-free sample would be very small. An estimate using the two-band model gives $<0.01$ carriers per unit cell. This implies a strong rearrangement of the Fe 4$s$ and Rh 4$d$ and 5$s$ bands, as has previously been predicted~\cite{Koenig:82}. It has recently become clear that in the case of weak hybridisation between a strongly-correlated band (in this case the Fe $3d$ band) and other (conduction) bands can lead to a \emph{selective} Mott transition~\cite{Koga:04,Pepin:07, deLeo:08}. In this scenario the localisation of Fe $3d$ electrons in FeRh due to the Mott mechanism is unaffected by the presence of other bands crossing the Fermi energy; defect bands arising from substitution defects or Fe $4s$ and Rh $4d$ and $5s$ bands. An orbital-selective Mott transition has for example been suggested to occur in the heavy-fermion intermetallic compound YbRh$_2$Si$_2$~\cite{Paschen:04}; at low temperatures the Yb $4f$ moments order antiferromagnetically, and in Hall effect measurements on this material a large discontinuity in the carrier density is observed at the field-induced Kondo-breakdown quantum-critical point, beyond which the ``Anderson impurity'' $4f$ moments are screened by the conduction electrons.

An orbital-selective Mott transition at the AF-FM transition in FeRh, selective to the Fe $3d$ orbitals, could also provide a natural explanation for the confluence of $n_n = n_p$ and $\mu _n = - \mu _p$ observed above the transition. The situation is then that the transport is dominated by a single half-filled conduction band with hopping-like conductivity for  both the electrons and the holes - an exciton gas~\cite{Kellogg:04}. Mott~\cite{Mott:61} suggested that this is the high-temperature metallic state in a metal-insulator transition. However, as is clear from the Ni-doped FeRh data~\cite{Kobayashi:01}, which does not show the exciton gas-like behaviour, the vanishing $R_{\rm{H}}$ in the high-temperature Hall data in FeRh may not be uniquely linked to the  AF-FM transition.

Finally, the (selective) Mott mechanism would provide a natural mechanism for the change of magnetic exchange between the Fe $3d$ moments. Antiferromagnetism is favoured for the localized Fe $3d$ electrons in the low-temperature phase, while for delocalized electrons Coulomb repulsions between the electrons can give rise to ferromagnetism by the Stoner mechanism~\cite{Fazekas:99}, which is responsible for this type of order in the $3d$ ferromagnets such as Fe.

\section{Conclusion}
In conclusion, comprehensive Hall effect measurements on $\alpha^\prime$-FeRh reveal a dramatic reduction in carrier density to a sub-metallic value on cooling through the metamagnetic transition into the AF phase. The drop is over an order of magnitude, meaning that whole bands could be disappearing, rather than that a fraction of a Fermi surface being gapped out. Furthermore, the carrier density in the low-temperature AF phase is found to be dominated by a defect band related to Fe/Rh substitution defects. These observations establish that the metamagnetic transition in FeRh is driven by an electronic transition. One possibility, consistent with all the experimental features we observe, is that the electronic transition is in fact a Mott transition selective to strongly-correlated Fe $3d$ electrons.

\ack
We gratefully acknowledge the help and advice from M. Ali, J.S. Weaver, and N.A. Porter in Leeds and S. Langridge at ISIS. This work is supported by the NSF grant DMR-0907007 and a linked grant through the materials world network scheme by the National Science Foundation under Grant No. DMR-0908767 and the UK Engineering and Physical Sciences Research Council, Grant No. EP/G065640/1.

%
%\bibliographystyle{iopart-num}
%\bibliography{my-bibliography}

\providecommand{\newblock}{}

\end{document}